\documentclass[prl,a4paper,superscriptaddress,twocolumn,showpacs,amsmath,amssymb,floatfix]{revtex4}
\pdfoutput=1

\usepackage{graphicx}
\usepackage{amssymb}
\usepackage{float}
\usepackage{hyperref}
 
\usepackage{color}

\begin{document}

\title{Thermodynamic description of world GDP distribution\\
over countries}

\author{Klaus M. Frahm}
\affiliation{\mbox{Univ Toulouse, CNRS, Laboratoire de Physique Th\'eorique, 
Toulouse, France}}
\author{Dima L. Shepelyansky}
%\homepage[]{http://www.quantware.ups-tlse.fr}
\affiliation{\mbox{Univ Toulouse, CNRS, Laboratoire de Physique Th\'eorique, 
Toulouse, France}}
%\date{today}
\date{December 6, 2025; Revised: XXXX}

\begin{abstract}
We apply the concept of Rayleigh-Jeans thermalization of classical fields
for a description of the world Gross Domestic Product (GDP) distribution
over countries. The thermalization appears due to
a variety of interactions between countries
with conservation of two integrals being total GDP and
probability (norm). In such a case there is an emergence of Rayleigh-Jeans
condensation at states with low GDP. This phenomenon has been studied
theoretically and experimentally in multimode optical fibers
and we argue that it is at the origin of emergence of
poverty and oligarchic phases for GDP of countries.
A similar phenomenon has been discussed recently in the framework 
of the Wealth Thermalization Hypothesis to explain the high inequality
of wealth distribution in human society and
companies at Stock Exchange markets. We show that the 
Rayleigh-Jeans thermalization  well describes
the GDP distribution during the last 50 years.
\end{abstract}

%\pacs{05.45.Mt, 67.85.Hj,  47.27.-i, 72.15.Rn}
%05.45.-a Nonlinear dynamics and chaos
%67.85.Hj 	Bose-Einstein condensates in optical potentials
%47.27.-i 	Turbulent flows
%72.15.Rn 	Localization effects (Anderson or weak localization) 
%
%47.35.-i 	Hydrodynamic waves
%47.35.Bb 	Gravity waves 
%89.75.-k 	Complex systems 

\maketitle

{\it Introduction.} Gross Domestic Product (GDP) is a monetary measure of
the total market value of all final goods and services produced by a country
during a year \cite{gdpwiki}. International organizations
as United Nations (UN), International Monetary Fund (IMF)
and World Bank (WB) yearly report GDP values for
world countries and territories as e.g in \cite{gdpcountries}
for recent years (with up to $N=212$ countries).
GDP country values for years 1970-2023
are publicly available from the UN statistics division at \cite{unyears}
and we mainly use this data in our work. If we consider countries as
independent players with equal rights according to the UN convention
then a most striking feature of the GDP distribution of countries 
is a strong inequality
when 60\% of all countries own only 2\% of the total GDP,
while 10\% (1\%) of countries own 82\% (44\%) of the total GDP.
(e.g. in 2023 \cite{gdpcountries,unyears}).
Of course, countries have quite different populations but
this argument does not work completely since the most populated countries as
India and China are not the owners of the biggest GDP.

It is interesting to note that this inequality of GDP distribution is
very similar to inequality of wealth distribution
in the human society (see e.g. \cite{piketty1,piketty2,boston}) where 
50\% of the world population owns only 2\%
of total wealth,
while 10\% (1\%) of population owns 75\% (38\%) of the total wealth
\cite{piketty2}.

Wealth distributions  are usually  characterized by the Lorenz curve
\cite{lorenz,boston} which depicts the dependence of cumulated 
normalized wealth $0\leq w \leq 1$ on the cumulated normalized fraction of
population or households $0 \leq h \leq 1$.
The limit of perfect wealth equipartition corresponds to
the diagonal $w=h$. The doubled area between diagonal
and the Lorenz curve $w(h)$ is called the Gini coefficient
$0 \leq G \leq 1$  \cite{gini,boston}. Values of $G \ll 1$ 
($G$ close to $1$) correspond to small (strong) inequality.
Values of $G$ can be found in \cite{wikigini}
for world countries in 2021 being in the range
$0.59 < G < 0.90$; for the whole world $G = 0.889$.
We keep in this work the notations of $w, h$
for cumulated country GDP (wealth like) and
cumulated number of countries (households like)
respectively to keep parallels with wealth inequality
studies.

In \cite{wth}, the Wealth Thermalization Hypothesis (WTH)
was proposed according to which the wealth shared in a
country or the whole world is described
by the Rayleigh-Jeans (RJ) thermal distribution \cite{landau,zakharovbook}
with two conserved quantities of system wealth and norm or number of agents. 
Moreover in \cite{wth}
it was shown that this approach also depicts well the Lorenz curves for
market capitalization (Market Cap) of companies
at Stock Exchange (SE) of Hong Kong, London and New York. 
The RJ thermal condensation appears also at long times  in systems of
dynamical oscillators
with a certain type of nonlinear perturbation that was demonstrated in 
\cite{wth} for  examples of social networks. 

At sufficiently low temperatures (low energy/global system wealth) the 
two integrals of motion lead naturally to RJ condensation where most of the 
probability is concentrated on one or a few low energy modes 
providing the physical mechanism for the appearance of a huge poor fraction
of population in countries or companies at SE \cite{wth}. 
At the same time a small oligarchic fraction of
households/companies captures a huge amount of the total wealth. 
The phenomenon of RJ condensation is not broadly known
in physics but it was theoretically and numerically
studied for multimode optical fibers \cite{picozzi1,picozzi2,ourfiber}.
It is also known in fibers as self-cleaning and 
was observed in multimode fiber experiments \cite{wabnitz,picozzi3}
where properties of RJ thermalization were studied experimentally
\cite{babin,chris,picozzi4}.

%%%%%%%%%%%%%%%%%%
%%% I think this part is not necessary ... It will be explained late
%Thus SE companies play a role of individual gamers 
%and their interactions lead to their RJ thermalization and condensation
%that generates a huge inequality. It is important to note
%that the number of workers in a given company
%does not appear in such an RJ description.
%This gives certain grounds that the RJ thermalization
%of GDP of countries is also not significantly influenced by country
%population. We discuss this point in detail at a later stage.

We should note that a different thermodynamic Boltzmann-Gibbs approach
to the wealth distribution was previously discussed 
in \cite{yakovenko1,yakovenko2,yakovenko3}. However, this approach 
does not capture the emergence of a huge condensate of poverty. 
The importance of two integrals of motion
has also been pointed out in the framework of 
a specific model of wealth evolution \cite{boghosian1,boghosian2}.
A variety of different models for the description of wealth inequality
has been proposed by different groups 
(see e.g. \cite{redner,bouchaud,chakraborti}) 
indicating that a more universal description is needed 
to explain the generic  phenomenon of inequality
existing in various countries.

{\it RJ thermalization and condensation for GDP.}
According to WTH \cite{wth} the wealth of a country population (or country GDP)
is described by the RJ thermal distribution \cite{landau,zakharovbook}:
\begin{equation}
\rho_m = \frac{T}{E_m-\mu} \; ({\rm RJ}) .
\label{eqrj}
\end{equation}
Here it is assumed that the system wealth
is distributed over certain states (assets) 
$0 \leq m < N$ with energies $E_m$ (representing the wealth value 
of asset $m$) 
and the population probabilities of these states are $\rho_m$ 
(population fraction owning asset $m$);
$T$ and $\mu$ are system temperature and chemical potential.
The conservation of two integrals of motion (norm and energy) implies
that $\sum_m \rho_m =1$ (we fix the norm to unity) and that 
the system average wealth (or total world GDP), 
being its average energy, is $ \sum_m E_m \rho_m =E$. 
The energy $E$ is a free parameter of the model and in the following we use 
the rescaled energy $\varepsilon=E/B$ where $B=E_{N-1}$ is the total 
energy bandwidth and we assume $0=E_0<E_1<E_2<\ldots<E_{N-1}=B$. 
The system temperature $T(E)$ and its chemical potential $\mu(T)$ 
are determined by two implicit equations for norm and average energy using 
(\ref{eqrj}). 
In this model, Lorenz curves are obtained from the set of points $(h(m),w(m))$ 
(for $m=0,\ldots,N$) with partial sums 
$h(m)=\sum_{l<m}\rho_m$ and $w(m)=\sum_{l<m} (E_m/E)\rho_m$ 
such that $h(0)=w(0)=0$ and $h(N)=w(N)=1$. 

The RJ distribution (\ref{eqrj}) follows also from the Bose-Einstein 
thermal distribution for quantum states 
at high temperature  with $T \gg  E_m$ \cite{landau}.
However, we assume that the wealth dynamics is described by
the classical field equations
so that RJ distribution is valid for any temperatures $T$
including negative ones \cite{wth,rmtprl}. 
It was shown \cite{wth,ourfiber,rmtprl} 
that a nonlinear perturbation of a generic system of linear oscillators
leads to a chaotic dynamics (for a perturbation above a chaos border)
with dynamical thermalization described by the RJ distribution over energies
(or oscillator frequencies $\omega_m \propto E_m$). 

The thermalization can have a dynamical origin
with a chaotic nonlinear dynamics leading to (\ref{eqrj})
or it can appear due to an external thermal bath.
We assume that a dynamical origin of WTH is more
adequate since in a first approximation on a scale of one year a country 
or the whole world can be considered to be quasi-isolated from slow 
external processes 
with approximate conservation of norm and energy.
We note that the GDP is changing from one year to another
but this change is small and can be considered as adiabatic
justifying the conservation of the two integrals of motion. 

To compare RJ results with real Lorenz curves as in \cite{wth} we consider two
models for the energies $E_m$. 
The first one is the simplest assumption that
density of state $\nu(E_m) = dm/dE_m$ is constant and
energies are equidistant ($0 \leq E_m = Bm/(N-1)$) where $N$ 
is the total number of states and $B=E_{N-1}$ is the energy bandwidth. 
We call this case the RJ standard (RJS) model. 
Its properties are discussed in detail in \cite{wth,ourfiber}.
The Lorenz curves, dependencies $T(\varepsilon)$, $\mu(\varepsilon)$,
phase diagram at various values of the dimensionless parameter 
$\varepsilon=E/B$ are shown in Figs.~I.1, I.2, A1, A2 in \cite{wth}. 
At fixed values of $\varepsilon$ the Lorenz curves do not depend on the 
particular choice of the global energy scale $B$.
As in \cite{wth}, 
to compare the RJS model with the real Lorenz curves we fix the 
parameter $\varepsilon$
by the condition that both cases have the same Gini coefficient $G$.
For the theoretical Lorenz curve based on (\ref{eqrj}) we use $N=10^4$ 
(the theoretical Lorenz curves are very similar between different values 
of $N$ as long as $N$ is sufficiently large). 

Following \cite{wth}, we also consider a more realistic model with 
$E_m = C[\exp(a m/N) - 1]/a$ and $\nu(E_m) = N/(1+aE_m/C)$.
It takes into account that there is a smaller number of states at
high revenues. Here $C\approx Ba/[\exp(a)-1]$ is a global constant related 
to the bandwidth $B$ whose value is not important for the particular 
form of the obtained Lorenz curves. We call this case the RJ exponential 
(RJE) model which has two parameters $a$ and $\varepsilon$. The limit 
$a\to 0$ corresponds 
to the RJS model with $E_m=Bm/N$. 
For a given value of the parameter $a$ the value of $\varepsilon(a)$ is 
determined by matching the Gini coefficient $G$ with its value from a real 
Lorenz curve and then the optimal value $a$ is computed by minimizing 
the average curve distance between the real Lorenz curve and the 
RJE Lorenz curve. 
In this way, we find that the RJE model gives results being very close 
to the real Lorenz curves 
(but at a price of two fit parameters $\varepsilon$, $a$ 
as compared to the RJS case having only one parameter $\varepsilon$). 
These RJS and RJE models
have growing state (asset) values $E_m$
that depict 
the social stratification phenomenon broadly
studied in sociology and 
known to be present in
a society \cite{marx,lenski,kerbo}.

\begin{figure}[t]
\begin{center}
\includegraphics[width=0.42\textwidth]{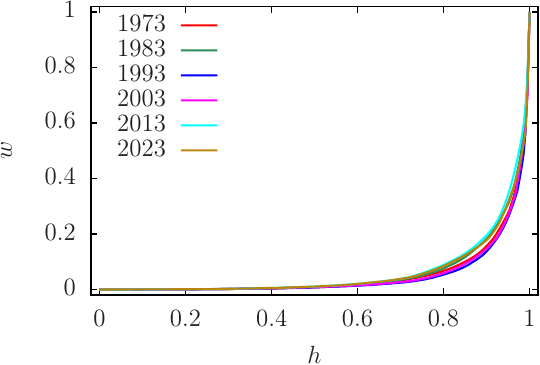}
\end{center}
\vglue -0.3cm
\caption{\label{fig1}
Lorenz curves of GDP for countries from UN data \cite{unyears} 
for 6 years between 1973 and 2023. The $x$-axis corresponds the 
cumulated fraction of households/countries ($h$) and the $y$-axis to
the cumulated fraction of wealth/GDP ($w$). 
}
\end{figure}

{\it GDP results}. The GDP Lorenz curves for years 1973, 1983, 1993, 2003, 2013, 2023,
obtained from UN data  \cite{unyears}, are shown in Fig.~\ref{fig1}
(a zoomed version is available at Supplementary Material (SupMat) Fig.~S1). 
In this time range of 50 years
the Lorenz curve remains remarkably stable showing only small variations from year to year
even if the total world GDP is changed from $5.23\times 10^{12}$  in 1973 to 
$1.05\times 10^{14}$ USD in 2023 
%100834 billions USD in 2023
and the total number of countries and territories  is changed from $N=187$ to 
$N=212$ during this period
(the list of all countries is available at \cite{gdpcountries,unyears}).
%and at related web page \cite{ourwebpage}).
The stability of the Lorenz curve confirms an adiabatic variation 
of the GDP world evolution and the approximate conservation of the 
two integrals of motion. 

\begin{figure}[t]
\begin{center}
  \includegraphics[width=0.42\textwidth]{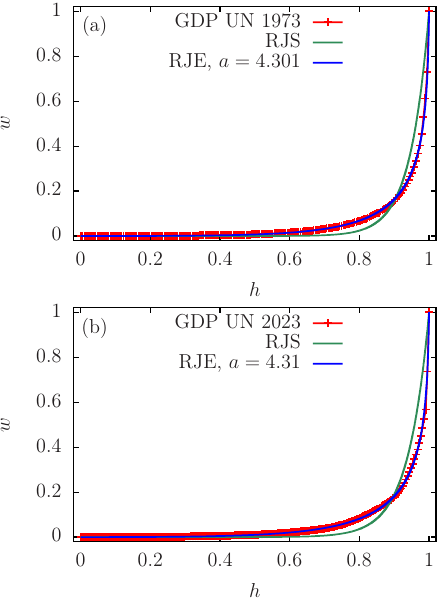}\\
\end{center}
\vglue -0.3cm
\caption{\label{fig2}
  Top panel (a): Lorenz curve for the year 1973 from UN data \cite{unyears}
  shown by red curve with (+), Gini coefficient is $G=0.892$;
  WTH theory with RJS model with $\varepsilon(RJS) = 0.0538$
  at same $G$ (green curve); RJE model results are shown with blue curve
  at $\varepsilon(RJE) = 0.00888$, $a=4.301$; bottom panel (b) shows same style data
  for the year 2023 with $G=0.88$, $\varepsilon(RJS) = 0.0601$ for the 
RJS model
and  $\varepsilon(RJE) = 0.01$, $a=4.31$ for the RJE model. Parameters
$T$, $\mu$ for RJS and RJE models for year 2023 at $N=212$
are given in SupMat.
}
\end{figure}

In Fig.~\ref{fig2}, we compare the GDP Lorenz curves of 1973 and 2023 with 
the WTH theory within the RJS and RJE models 
(the comparison for years 1983, 1993, 2003, 2013
is given in SupMat  Fig.~S2). The Lorenz curves for 1973
and 2023 years are rather close to each other
and thus the parameters of RJS and RJE models are also close
for these 2 years (same is valid for the other 4 years
shown in SupMat Fig.~S2). On a scale of 50 years
the system parameters have only modest variations 
in the ranges:
$0.871 \leq G \leq 0.904$; $0.0479 \leq \varepsilon(RJS) \leq 0.0646$;
$0.0078 \leq \varepsilon(RJE) \leq 0.0157$; $3.622 \leq  a \leq 4.31$. 
The RJS model describes well the global 
behavior of Lorenz curves but has visible deviations while the RJE model 
gives almost perfect agreement with real data. 

Of course one can express a criticism saying that
a fit of a monotonic curve with two parameters (as RJE case)
may give a rather good agreement but only the parameter $a$ 
influences the form of the spectrum $E_m$ and 
the other parameter $\varepsilon$ is simply the 
rescaled energy. Furthermore, 
in our opinion the most important point is not the almost perfect
agreement of two curves of real data and the RJE model
(even if it is useful to have it)
but the physical origin of RJ condensation
that naturally explains the appearance of the 
phase of high poverty and the oligarchic phase. 
Thus for the year 2023 60\% of countries own 2\% of the total GDP and 
10\% (1\%) of countries own 82\% (44\%) of GDP 
(values obtained by linear interpolation from the raw data). 
These numbers are rather similar to those of wealth inequality for
the whole world \cite{piketty2}.
The physical reason of RJ condensation is a small rescaled value of total
system energy $\varepsilon = E/B \sim 0.01 \ll 1$.

\begin{figure}[t]
\begin{center}
\includegraphics[width=0.48\textwidth]{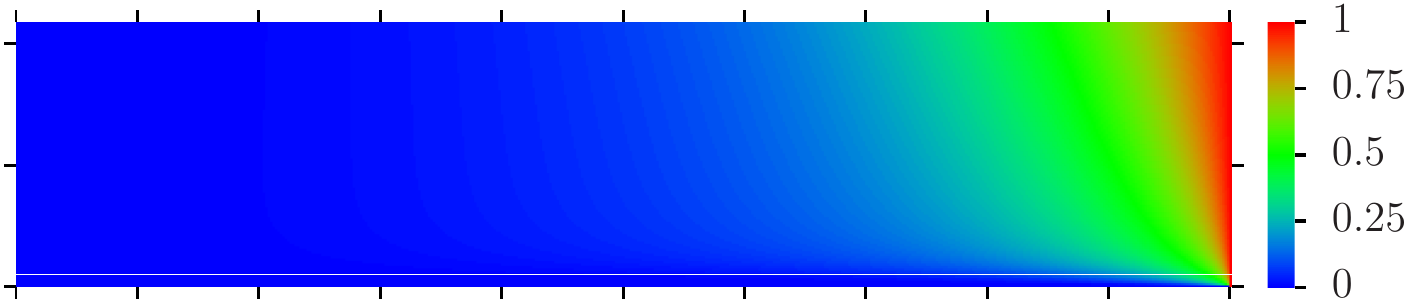}
\end{center}
\vglue -0.3cm
\caption{\label{fig3}
Color plot of wealth $w$ from Lorenz curves for the from the RJE model 
at $a=4.31$. The $x$-axis corresponds to 
the fraction of households $h\in[0, 1]$ 
and the $y$-axis to the rescaled energy  
$\varepsilon\in[0, \varepsilon_C[$ 
where $\varepsilon_C=0.218$ is the critical value 
at which the transition from $T>0$ to $T<0$ appears. 
The ticks mark integer multiples of 0.1 
for $h$ and $\varepsilon$. The white line corresponds to 
$\varepsilon(RJE)=0.01$ obtained from the fit of the RJE model 
using the GDP data of 2023 shown in Fig.~\ref{fig2}.
}
\end{figure}

It is possible to decrease the fraction of poverty phase by increasing
the dimensionless system energy $\varepsilon(RJE)$ as it is shown in 
Fig.~\ref{fig3} for RJE model at year 2023
(at the same time the structure of energies $E_m$ is fixed with 
$a=4.31= const$ 
being independent of $\varepsilon(RJE)$). Indeed, the
results of Fig.~\ref{fig3} show that an increase of $\varepsilon(RJE)$
gives a significant reduction of the poverty phase (shown by blue color)
however this requires a quite strong boost of this $\varepsilon(RJE)$
parameter that may not be an easy task.
We note that here, as in \cite{wth} for wealth of country households,
we consider only the cases with positive temperature
which for the RJE model at $a=4.31$ corresponds to 
 $\varepsilon(RJE) < \varepsilon_C=0.218$.

In addition to the WTH results based on UN data \cite{unyears}, 
we show the results obtained with GDP data sets from IMF (2025) and 
WB (2024) \cite{gdpcountries} (see SupMat Figs. S3 and S4). 
These IMF and WB results are very similar to those of UN.

\begin{figure}[t]
\begin{center}
\includegraphics[width=0.42\textwidth]{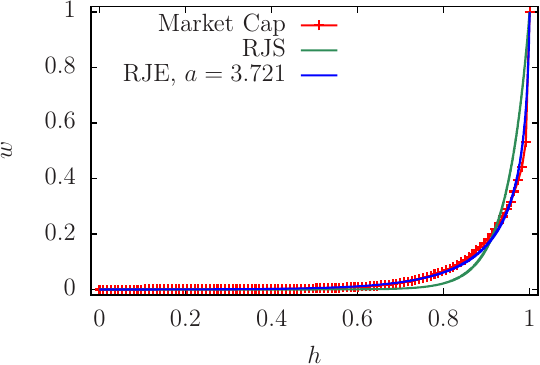}
\end{center}
\vglue -0.3cm
\caption{\label{fig4}
  The Lorenz curve for SE market capitalization of $N=113$ countries,
  data from \cite{marketcap}. The Lorenz curves are shown
  in the same style as in Fig.~\ref{fig2}. Here $G=0.895$,
  $\varepsilon(RJS) =0.0525$ for the RJS model,
  $\varepsilon(RJE) =0.0119$, $a=3.721$ for the RJE model.
}
\end{figure}

\begin{figure}[t]
\begin{center}
\includegraphics[width=0.48\textwidth]{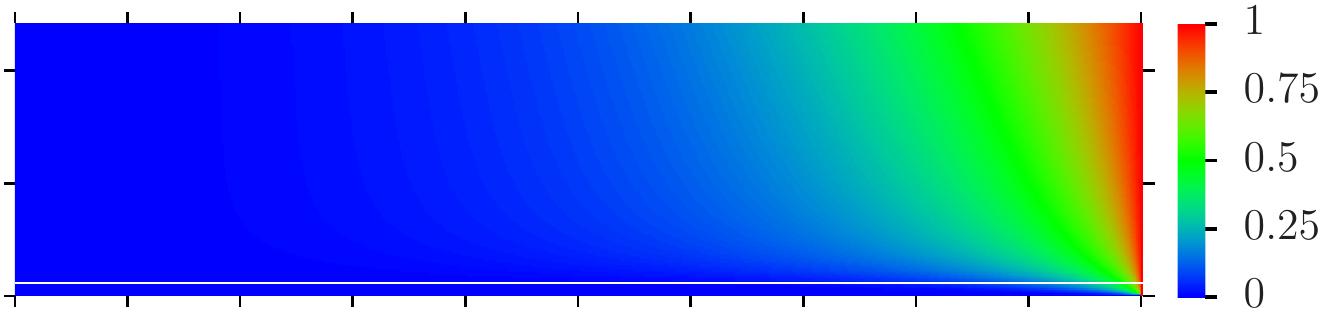}
\end{center}
\vglue -0.3cm
\caption{\label{fig5}
As Fig.~\ref{fig3} with $a=3.721$, $\varepsilon_C=0.244$ and 
a white line at 
$\varepsilon(RJE)=0.0119$ obtained from the fit of the RJE model 
using the SE Market Cap data shown in Fig.~\ref{fig4}.
}
\end{figure}

Another wealth measure of a country can be expressed
via the total SE market capitalization of all domestic companies 
listed in the country SE
(with World Bank data available at Wikipedia \cite{marketcap}).
In recent years 2024-2025 this list has $N=113$ countries
with the top total Market Cap $M=62186$ billion USD for USA
and minimal one $M=0.388$ million USD for Mongolia.
The ratio of Market Cap to GDP changes from country to country
being about 2.1 and 0.6 for USA and China, 0.8 for UK, 1.3 for France 
and 0.4 for Germany.
Thus the Market Cap represents a complementary country wealth measure 
in addition to the GDP. 
The Lorenz curves for Market Cap data \cite{marketcap}
are shown in Fig.~\ref{fig4} in the same presentation style 
as in Fig.~\ref{fig2}
for GDP. As for GDP data we find that the WTH describes well
the real data: certain deviations are present for the RJS model
while the RJE model describes the data almost perfectly.
The parameters of RJS and RJE models are close to those
of the GDP cases in Fig.~\ref{fig2}.
The poverty phase that owns 2\% of total Market Cap 
corresponds to 68\% of countries, 10\% (1\%) of countries own 82\% 
(48\%) of the total Market Cap.
For the RJE model the variation of the poverty and oligarchic
phase with $\varepsilon(RJE)$ (for $a=3.721$, $\varepsilon_C=0.244$) 
are shown in Fig.~\ref{fig5} which is similar to the GDP case in 
Fig.~\ref{fig3}. 

The WTH approach describes well the GDP distributions, wealth 
of country households \cite{piketty2}, and distributions of 
Market Cap of SE companies  presented in \cite{wth}. 
Thus for three types of players
the WTH theory explains very well the
existing inequality: for wealth of individual persons,
market capitalization of companies and GDP distribution over countries.
The number of persons for each of such a player
is very different. But we argue that
the thermalization takes place for an  individual player
and not for individual persons that belong to a given player.
This situation is similar to a case of gas in a 3D box composed of 
different atoms: even if the masses are different
for different sorts of atoms still the
average  kinetic energies of atoms are the same
being equal to $3kT/2$ where $k$ in the Boltzmann constant
and $T$ is temperature. However, the atom square velocity is
inversely proportional to atomic mass.
Similarly, for the GDP per capita \cite{capita}
the three top positions correspond to 
Monaco, Liechtenstein, Luxembourg
that do not produce a significant influence on the world economy.
Of course, for GDP of countries
certain effects of high or small country population
may play some role but we argue that this does not
affect the RJ thermalization of the GDP distribution.
For completeness we present also the Lorenz curve
of population of countries, taken from \cite{popula},
(see SupMat Fig. S5) which also has a shape
similar to those of RJS and RJE models,
but we do not discuss possibilities of
thermalization of population of countries.

{\it Discussion.} The obtained results show that the Lorenz curve of world 
GDP distribution remains stable during last 50 years with only weak variations 
from year to year. This stresses the universal foundations being
at the grounds of this distribution.
We proposed here the WTH theory
according to which GDP distribution is described by the RJ thermodynamic 
distribution that appears as a result of various interactions 
between countries,
e.g. international trade. In this RJ approach there is a phenomenon of
RJ condensation leading to concentration a huge fraction of countries
at very low GDP values (e.g. 60\% own only 2\% of total GDP).
The RJ description assumes the existence of certain thermal states
with energy spectrum $E_m$ describing levels of GDP. 
For the simplest case with a constant density of these states
we obtain the RJS model that describes well the real Lorenz curves
but deviations from them are still present.
The extended RJE model, where the density of states decreases at high 
GDP values, 
gives almost perfect description of real Lorenz curves.
However, for us the main argument in the favor
of WTH description of GDP distribution
is that the phenomenon of RJ condensation
gives a fundamental explanation of the 
GDP inequality. Moreover, the WTH theory \cite{wth}
also describes the inequality of country households \cite{piketty2}
 and inequality 
of market capitalization of companies at
SE of Hong Kong, London, New York \cite{wth}.

We hope that the WTH description of inequality
in different society groups
will find further useful
applications. 

\noindent {\bf Acknowledgments:}
This work has been partially supported through the grant
NANOX $N^o$ ANR-17-EURE-0009 in the framework of 
the Programme Investissements d'Avenir (project MTDINA).

%%%%%%%%%%%%%%%%%%%%%%%%%%%%%%%%%%%%%%%%%%%%%%%%%%%%%%%%%

%\documentclass[prl,a4paper,superscriptaddress,twocolumn,showpacs,amsmath,amssymb,floatfix]{revtex4}
%\documentclass[pra,a4paper,superscriptaddress,twocolumn,amsmath,amssymb,floatfix]{revtex4}
%\documentclass[pra,a4paper,superscriptaddress,twocolumn,showpacs,amsmath,amssymb,floatfix]{revtex4}

%\usepackage[dvips]{graphicx}
%\usepackage{graphicx}
%\usepackage{amsmath}
%\usepackage{amssymb}
%\usepackage{float}
%\usepackage{hyperref}

%\usepackage{color}
\def\eps{\varepsilon}
\def\folgt{\quad\Rightarrow\quad}

%\textheight=21cm

%\input{epsf}

%\begin{document}
%\title{\phantom{dummy}}
%\title{Supplementary Material for Wealth Thermalization Hypothesis}

%\maketitle

\newpage

\setcounter{figure}{0} \renewcommand{\thefigure}{S\arabic{figure}} 
\setcounter{equation}{0} \renewcommand{\theequation}{S.\arabic{equation}} 
\setcounter{page}{1}

%%% extra part for manual adding of section numbers
%%% use \mysection{...} instead of \section{...}
\newcounter{mysec}
\setcounter{mysec}{0}
\newcommand{\mysection}[1]{
\stepcounter{mysec}
\section{\Roman{mysec}. #1}
}

\noindent{{\bf Supplementary Material for\\
\vskip 0.2cm
\noindent{Thermodynamic description of world\\
  GDP distribution over countries}\\}
\bigskip

\noindent by
K.~M.~Frahm and D.~L.~Shepelyansky\\
\noindent Univ Toulouse, CNRS. Laboratoire de Physique Th\'eorique, 
 Toulouse, France
\bigskip

Submitted December 6, 2025 with  
additional figures and explanations.

Here we present additional Figures as a complement
to the main part of the paper.

\begin{figure}[h]
\begin{center}
\includegraphics[width=0.42\textwidth]{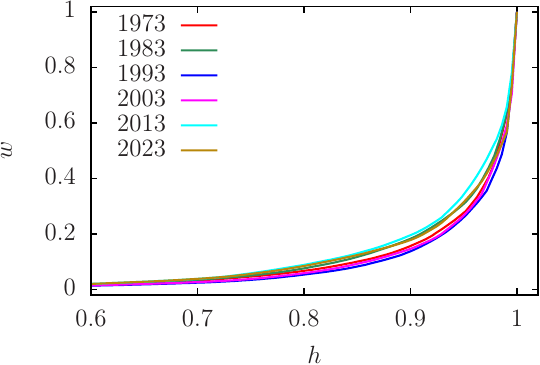}%
\end{center}
\caption{\label{figS1}
Zoom of Fig. 1 for $h\in[0.6,1]$.
}
\end{figure}

\begin{figure}[h]
\begin{center}
\includegraphics[width=0.42\textwidth]{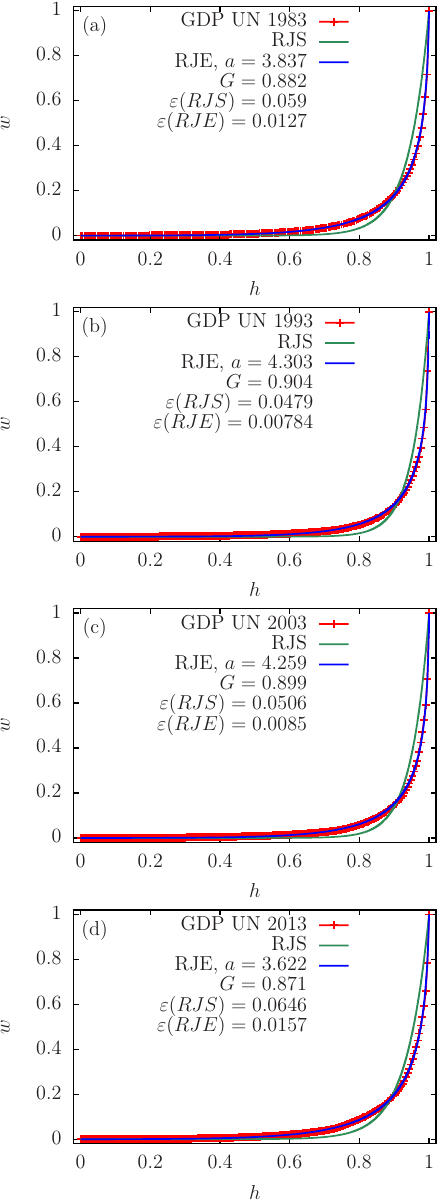}%
\end{center}
\caption{\label{figS2}
\label{fig_rhodensbar}
As Fig. 2 using the GDP UN data of Ref. [3] for the years 1983 (a), 
1993 (b), 2003 (c) and 2013 (d). 
}
\end{figure}

We also give the approximate values of system
parameters $B, T, \mu$ for the two cases 
of the RJS and the RJE model for the UN data of 2023 year in 
Fig.~2(b). 
For this year we have $N=212$ countries and territories
with the total world GDP being $E \approx 100$ trillion USD $=10^{14}$ USD.
For  the RJS model at $\varepsilon(RJS) =E/B =0.0601$ of Fig.~2b this
gives $B =E/\varepsilon \approx 1670$ trillion USD
with a spacing between levels $\Delta(RJS) = E_{m+1} -E_m =B/N  \approx 7.9$ 
trillion USD.
The solution of the two equations for the two conserved integrals:
$\sum^{N-1}_{m=0} E_m \rho_m =E$ and
$\sum^{N-1}_{m=0} \rho_m=1$ with $\rho_m=T/(E_m-\mu)=(E-\mu)/[N(E_m-\mu)]$ 
gives $\mu = -0.74$ trillion USD, 
$T= (E-\mu)/N\approx E/N\approx 0.47$ trillion USD $\ll \Delta $
and the probability of the ground state 
is $\rho_0 = -T/\mu\approx 0.64$ indicating a rather strong RJ 
condensation. \\
For the RJE model we have 
$\mu=-1.57$ trillion USD, $T\approx E/N\approx 0.47$ trillion USD and 
$\rho_0=-T/\mu \approx 0.30$. Here the initial level spacing 
(at $m\ll N=212$ and using $a=4.31$) 
is $\Delta(RJE)=C/N=Ba/[N(e^a-1)]\approx B(RJE)/(17N)=E/(17 N \eps(RJE))
\approx (6/17)\Delta(RJS)\approx 2.8$ trillion USD which is still 
larger than $T$ but smaller then $\Delta(RJS)$. 
%%% optionnel explication below (can be removed)
Despite of the smaller value of $\varepsilon(RJE)=0.01$ the effect of 
RJ condensation for the RJE model in comparison to the RJS case is a 
bit reduced, but still clearly present since the effect of the reduced 
value of $\varepsilon$ (gives a factor $6$ for $\Delta$) is 
more than compensated by the the factor $1/17$ from the exponential 
factor $a/(e^a-1)\approx 1/17$ due to the RJE model.

\begin{figure}[h]
\begin{center}
\includegraphics[width=0.42\textwidth]{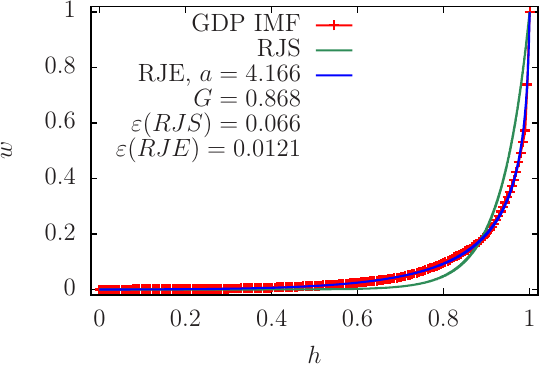}%
\end{center}
\caption{\label{figS3}
As Fig. 2 using the GDP IMF data of Ref. [2] for the year 2025. 
}
\end{figure}

\begin{figure}[h]
\begin{center}
\includegraphics[width=0.42\textwidth]{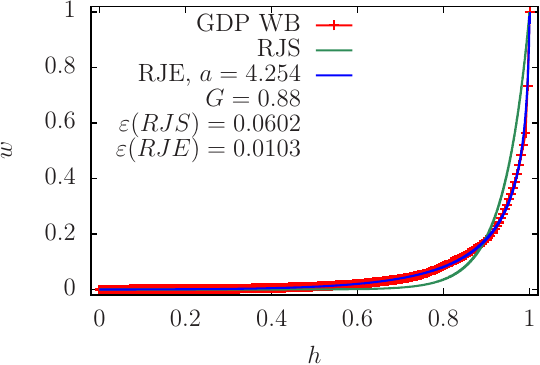}%
\end{center}
\caption{\label{figS4}
As Fig. 2 using the GDP World Bank data of Ref. [2] for the year 2025. 
}
\end{figure}

\begin{figure}[h]
\begin{center}
\includegraphics[width=0.42\textwidth]{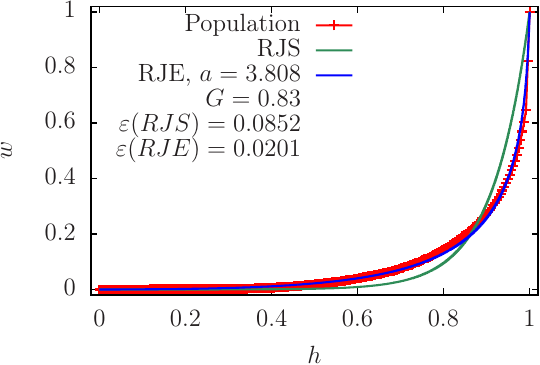}%
\end{center}
\caption{\label{figS5}
As Fig. 2 but for the population of countries
for the year 2023 using the UN data of Ref. [32]. 
Here the Lorenz curve is computed assuming that each country represents 
the same household fraction. 
}
\end{figure}

\end{document}